\pdfoutput=1
\documentclass[sigconf,nonacm]{acmart}
\AtBeginDocument{}

\usepackage{booktabs}
\usepackage{float}
\usepackage{enumitem}
\usepackage{xcolor}
\usepackage{tikz}
\usetikzlibrary{arrows.meta,bending,positioning,calc,shapes.geometric,fit}
\usepackage{cleveref}
\usepackage{acro}
\acsetup{use-only-first=true}
\DeclareAcronym{AMM}{short=AMM, long=automated market maker}
\DeclareAcronym{CFMM}{short=CFMM, long=constant-function market maker}
\DeclareAcronym{DEX}{short=DEX, long=decentralized exchange}
\DeclareAcronym{CEX}{short=CEX, long=centralized exchange}
\DeclareAcronym{DeFi}{short=DeFi, long=decentralized finance}
\DeclareAcronym{TVL}{short=TVL, long=total value locked}
\DeclareAcronym{FDV}{short=FDV, long=fully diluted valuation}
\DeclareAcronym{BSC}{short=BSC, long=BNB Smart Chain}
\DeclareAcronym{CMC}{short=CMC, long=CoinMarketCap}
\DeclareAcronym{EVM}{short=EVM, long=Ethereum Virtual Machine}
\DeclareAcronym{RPC}{short=RPC, long=remote procedure call}
\DeclareAcronym{API}{short=API, long=application programming interface}
\DeclareAcronym{ETF}{short=ETF, long=exchange-traded fund}
\DeclareAcronym{GFV}{short=GFV, long=generalized flow valuation}
\DeclareAcronym{LP}{short=LP, long=linear program}
\DeclareAcronym{bp}{short=bp, long=basis point, short-plural=}
\DeclareAcronym{LMDI}{short=LMDI, long=logarithmic mean Divisia index}
\DeclareAcronym{SE}{short=SE, long=standard error}
\DeclareAcronym{CI}{short=CI, long=confidence interval}
\DeclareAcronym{OLS}{short=OLS, long=ordinary least squares}
\DeclareAcronym{HAC}{short=HAC, long=heteroskedasticity and autocorrelation consistent, short-plural=}
\DeclareAcronym{VAR}{short=VAR, long=vector autoregression}
\crefname{appendix}{Appendix}{Appendices}\crefname{subappendix}{Appendix}{Appendices}\crefname{subsubappendix}{Appendix}{Appendices}
\crefname{proposition}{Proposition}{Propositions}\crefname{lemma}{Lemma}{Lemmas}\crefname{definition}{Definition}{Definitions}\crefname{corollary}{Corollary}{Corollaries}\crefname{assumption}{Assumption}{Assumptions}\crefname{theorem}{Theorem}{Theorems}\crefname{example}{Example}{Examples}\crefname{remark}{Remark}{Remarks}
\graphicspath{{figures/}}

\AtBeginDocument{}

\newcommand{\RI}{risk import}
\newcommand{\ES}{exit sharing}
\newcommand{\VB}{value backflow}
\newcommand{\bI}{(I)}
\newcommand{\bII}{(II)}
\newcommand{\bIII}{(III)}
\colorlet{biasIcol}{red!70!black}
\colorlet{biasIIcol}{orange!85!black}
\colorlet{biasIIIcol}{blue!70!black}
\tikzset{biasicon/.style={x=0.5em,y=0.5em,baseline=-0.32em,line cap=round,line join=round}}
\DeclareRobustCommand{\iconI}{\tikz[biasicon]{\fill[biasIcol](0,0)circle(1);\draw[white,line width=0.6pt](-0.62,0) sin (-0.31,0.35) cos (0,0) sin (0.31,-0.35) cos (0.62,0);}}
\DeclareRobustCommand{\iconII}{\tikz[biasicon]{\fill[biasIIcol](0,0)circle(1);\draw[white,line width=0.6pt](-0.5,-0.55)--(0,-0.02)--(0.5,-0.55);\draw[white,line width=0.6pt,-{Stealth[length=0.22em,width=0.3em]}](0,-0.02)--(0,0.68);}}
\DeclareRobustCommand{\iconIII}{\tikz[biasicon]{\fill[biasIIIcol](0,0)circle(1);\draw[white,line width=0.6pt,-{Stealth[length=0.22em,width=0.3em,bend]}](-0.275,-0.476)arc[start angle=240,end angle=-60,radius=0.55];}}

\newcommand{\GFV}{\ensuremath{\mathrm{GFV}}}
\newcommand{\Cut}{\ensuremath{\mathrm{Cut}}}

\newcommand{\Mhub}[1]{\ensuremath{M^{(#1)}}}
\newcommand{\Mstd}{\ensuremath{M}}
\newcommand{\Tone}{\ensuremath{T_1}}
\newcommand{\Ttwo}{\ensuremath{T_2}}
\newcommand{\Tthree}{\ensuremath{T_3}}
\newcommand{\dlog}{\ensuremath{\Delta\!\log}}
\DeclareMathOperator{\Var}{Var}
\DeclareMathOperator{\Cov}{Cov}
\DeclareMathOperator{\down}{down}
\DeclareMathOperator{\up}{up}

\newcommand{\usd}[1]{\$#1}

\begin{document}

\title{The Pairing Bias: Pricing and Liquidity Distortions in Graph-Structured Markets}

\author{Weilin Li}
\affiliation{\institution{University College London}\city{London}\country{United Kingdom}}
\email{weilin.li.24@ucl.ac.uk}

\author{Jiahua Xu}
\affiliation{\institution{University College London}\city{London}\country{United Kingdom}}
\email{jiahua.xu@ucl.ac.uk}

\begin{abstract}
Unlike stock markets, where every asset is quoted in the local currency, on-chain AMM markets are graph-structured. A token can be paired with any other token, and its USD price is derived through intermediate tokens (its hubs, such as ETH). Market-data platforms and traders still rely on standard metrics such as market cap, which ignore how tokens are paired and thus distort measured value and risk. We call this distortion the \emph{pairing bias}, which takes three forms. (1)~Risk import: a token imports its hub's volatility. (2)~Exit sharing: tokens that share a hub compete for its liquidity when sold. (3)~Value backflow: buying a token raises its hub's price. Issuers and platforms, who choose the pairings, can exploit these biases. We quantify the pairing bias on Ethereum, Base and BSC. On Ethereum, 76.3\% of float market cap outside ETH, BTC and stablecoins is priced through a non-USD hub, and ETH's price accounts for 77.1\% of the daily variation in the market cap of ETH-paired tokens. When a hub's holders sell alongside its paired tokens, shared exits cut the tokens' proceeds by up to a further 44.3\%. On two platform hubs priced on chain, purchases of their paired tokens came with rises in the hubs' prices. We propose topology-aware valuation, which bounds realizable value and reports hub exposures.
\end{abstract}

\ccsdesc[500]{Information systems~Digital cash}
\ccsdesc[500]{Applied computing~Economics}
\ccsdesc[300]{Information systems~Data mining}

\keywords{Decentralized finance (DeFi), Automated market makers, Valuation, Market structure}

\maketitle

\begin{figure}[!t]
\centering
\includegraphics[width=0.96\columnwidth]{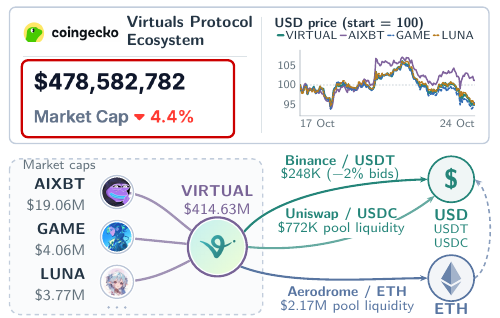}
\caption{CoinGecko reports one market cap for the Virtuals ecosystem, but the ecosystem's tokens are priced through VIRTUAL. Their prices move with VIRTUAL's, and the total market cap exit to USD through a few VIRTUAL markets.}
\label{fig:overview}
\end{figure}

\section{Introduction}\label{sec:intro}

Cryptocurrency traders read the market through a few numbers that Web platforms and wallets publish, such as each token's price, market cap and liquidity. CoinMarketCap and CoinGecko rank tokens by market cap and show their USD prices~\cite{coinmarketcap,coingecko}. CoinGecko also reports the summed market cap of each token ecosystem~\cite{CoinGeckoVirtualsCategory} (\Cref{fig:overview}). These \emph{standard metrics} come from traditional finance, and each multiplies an asset's USD price by a quantity and sums over assets. They presume a \emph{star market}, a topology in which every asset trades directly against USD in a separate order book, as in the US stock market. Trading one asset will not affect another, so each price reflects only its own market and values can be added up.

By comparison, on-chain \ac{AMM} markets have a \emph{graph-structured} topology. Anyone can pair any two tokens, and a token's USD price is the product of the prices along its route to USD, which can pass through several intermediate tokens (\emph{hubs}) such as ETH, BNB or VIRTUAL (\Cref{fig:overview}). On Ethereum, outside ETH, BTC and stablecoins, 76.3\% of float market cap is priced through at least one non-USD hub, and 56.3\% sits in tokens paired with at least two different tokens. A token's price and value then depend on more than its own market. Standard metrics, which read each token in isolation, therefore misread them, and we call these misreadings the \emph{pairing bias}. This paper asks where the pairing bias comes from, how large it is, and how to correct it.

The pairing bias takes three forms (\Cref{fig:mechanism}). First, a token's USD price moves with every hub on its route, so its holders bear the hubs' price risk. Standard metrics cannot separate these hub-driven moves from those driven by trading in the token itself, and we call this \iconI~\emph{risk import}. Second, the tokens paired with a hub, its \emph{leaves}, reach USD through the hub's markets, which the hub's own holders also use. When they sell together, they compete for this exit, and each realizes less. Standard metrics add up market caps and liquidity (\Cref{fig:overview}) as if each token had its own exit, and we call this \iconII~\emph{exit sharing}. Third, buying a token also buys every hub on its route, so demand for the token lifts its hubs' prices. Summed market caps count this demand twice, once in the token and once in its hubs. We call this \iconIII~\emph{value backflow}.

The pairing bias arises in any graph-structured market, including foreign exchange~\cite{Krugman1980Vehicle,Devereux2013Vehicle}. Japanese stocks reported in USD are priced through the yen, a stable and deep hub, so the bias is mild. On-chain hubs such as ETH and VIRTUAL are far more volatile and shallower, and sit in a more complex topology. The pairing bias also differs from illiquidity, which inflates market caps~\cite{Mongardini2025MemeManipulations} and would in a star market too. Illiquidity comes from a token's own thin market and is corrected token by token with a haircut. The pairing bias comes from its hubs' markets, so a token with a deep pool against its hub still imports the hub's risk, shares its exit and lifts its price. The pairing bias matters because these metrics decide how tokens are ranked, listed and traded. It can also be exploited, as issuers and platforms choose the pairings. We make three contributions.
\begin{enumerate}[leftmargin=*,itemsep=2pt,topsep=3pt]
 \item \textbf{A framework for the pairing bias.} We introduce the pairing bias and formalize it in a graph model of on-chain markets (\Cref{sec:model,sec:biases}). We decompose a token's price into the contributions of its hubs, bound the realizable value of tokens sharing an exit, and derive how a hub's price responds to demand for the tokens it prices. We also show that an issuer can inflate market cap without bound, using little external capital, by manipulating the pairing of its tokens.
 \item \textbf{Empirical measurement of the pairing bias.} We measure the pairing bias on Ethereum, Base and BSC (\Cref{sec:measurement}). ETH's price accounts for 77.1\% of the daily variation in the total market cap of ETH's leaves on Ethereum. When a hub's holders sell 1\% of its float alongside its leaves, the leaves lose up to a further 44.3\% of their standalone proceeds. On two platform hubs priced on chain, net leaf purchases worth 0.01\% of the hub's supply in an hour came with a 0.68--0.75\% rise in the hub's price.
 \item \textbf{Topology-aware valuation.} We propose metrics that keep the inputs of the standard metrics but account for the topology (\Cref{sec:gfv}). An exposure vector reports how much of a token's value passes through each hub, which shows the risk it imports. We also propose \ac{GFV}, which values a token set by a partial liquidation, solved as a convex routing problem. A simulated sale bounds it from below and the capacity of the shared exits bounds it from above. For value backflow, we recommend that platforms disclose what share of a hub's pooled supply sits in its leaves' pools (\Cref{sec:discussion}).
\end{enumerate}

\section{Background}\label{sec:background}

\textit{DEX and Route.} Unlike a \ac{CEX}, which matches orders in an off-chain order book, a \ac{DEX} settles every trade on chain and lets anyone create a market for any pair of tokens. The most common \ac{DEX} design is the \ac{AMM}, or \ac{CFMM}, where a pool holds reserves of two or more tokens and prices every trade by a formula of its reserves~\cite{Angeris2023TheMakers,Xu2023SoK:Protocols}. Examples are Uniswap v2, v3, v4 and the stableswap pools of Curve. Each trade moves the reserves along a convex curve, so a larger trade gets a worse average price.
A trade between two tokens without a common pool follows a \emph{route} of pools through intermediate tokens. Prices follow the same routes. A token is quoted in the token it is paired with, its \emph{hub}, and its USD price is synthesized along the route. For example, a token $i$ quoted in hub $h$ is priced at $p_{i\to \mathrm{USD}}=p_{i\to h}\,p_{h\to \mathrm{USD}}$, its local rate times the hub's USD price. A route can pass through several intermediate hubs. Routers search over routes and split a trade across pools. Finding the best route is a separate problem~\cite{Angeris2022OptimalMakers}.

\textit{Platforms and tokenized stocks.} On Base, the token-launch platform Virtuals quotes its agent tokens in VIRTUAL, and Zora quotes creator coins in ZORA and content coins in their creator's coin~\cite{VirtualsWhitepaper,ZoraCoinsDocs}. Some platforms also let creators quote in a \emph{tokenized stock}, a token whose issuer mints and redeems it for authorized parties to keep its on-chain price near the stock's.

\textit{Standard Metrics.} Market-data platforms and \ac{DeFi} protocols value tokens with a few standard metrics, which are price and price change on every platform (\iconI\,\iconIII), market cap and \ac{FDV} in rankings, listings and indices~\cite{coinmarketcap} (\iconI\,\iconII\,\iconIII), liquidity on \ac{DEX} screeners (\iconI\,\iconII\,\iconIII) and portfolio value in wallets and lending protocols (\iconI\,\iconII\,\iconIII). The icons mark the pairing biases of \Cref{sec:intro} that affect each metric. Each metric sums quantities valued at routed USD prices.

\section{Model}\label{sec:model}
\subsection{Graph-structured markets}\label{sec:market}

We model the market as a directed graph $G=(V,E)$ whose nodes are tokens and whose edges are the trading venues. We use US dollars as the \emph{anchor}, the unit in which prices and values are reported, as crypto markets do. Any reporting currency works the same way, such as yen for stocks listed in Japan. The \emph{sink} set $\mathcal U\subseteq V$ holds the USD stablecoins and a $\mathsf{usd}$ node, counted at \$1. A token \emph{exits} when it is converted into a sink. Token $i$'s circulating supply is $q_i$.

A \emph{venue} is a \ac{DEX} pool or an order book. A pool $P$ holds reserves $R_P$ over its coin set $C(P)\subseteq V$, $|C(P)|\ge2$. For coins $u\ne v$ of $P$, its \emph{exchange function} $\Psi^{uv}_P(\Delta;R_P)$ is the amount of $v$ it pays out for an input $\Delta\ge0$ of $u$ at the current reserves $R_P$. It is zero at zero, increasing and concave in $\Delta$ (price impact), and never exceeds the reserve $R_P(v)$. The \emph{price} of $u$ in units of $v$ is the marginal rate of $\Psi^{uv}_P$ at zero before fees (e.g., $R_P(v)/R_P(u)$ for Uniswap v2). Each pool contributes an edge $u\to v$ for every ordered pair of its coins. A \ac{CEX} book $b$ sells its token $v(b)$ for $\mathsf{usd}$ at its visible bids, up to their total value $D_b$, and contributes the edge $v(b)\to\mathsf{usd}$. A market is a \emph{star market} if every venue has at most one coin that is not a sink. Otherwise it is \emph{graph-structured} (such as AMM and FX markets).

A \emph{route} of token $i$ is a directed path in $G$ from $i$ to a sink, and the price is the product of the prices along the path. Standard metrics price each token along one route, its \emph{price route}, usually the one with the deepest liquidity, and this price is the token's \emph{mark} $p_i$. The price routes form a \emph{price forest} whose roots are sinks. The next token on the price route of $i$ is its \emph{quote asset}, and its \emph{local rate} $\ell_i$ is its price in units of its quote asset, with $\ell_i=1$ for a sink. The \emph{upstream nodes} $\up(i)$ are the non-sink tokens after $i$ on its price route, so $p_i=\ell_i\prod_{v\in\up(i)}\ell_v$, and the \emph{downstream set} of $h$ is $\down(h)=\{i:h\in\up(i)\}$. A token can be paired with several tokens and has many routes, which need not agree on a price. The price forest is only the price feed's choice of edges, and the market remains graph-structured, with trades on all edges.

A token with a nonempty downstream set is a \emph{hub}, and every other token that is not a sink is a \emph{leaf}. The leaves of a hub $h$ are the leaves in $\down(h)$. A \emph{designated hub} is a quote asset set by convention or platform rule, such as ETH, BNB, SOL, VIRTUAL, ZORA and AERO. Every other hub is a \emph{local hub}.

\subsection{The star market and standard metrics}\label{sec:standard}\label{sec:anchor}

A \emph{standard metric} is a linear aggregate of marks, $\sum_{v}x_v\,p_v$, in which the weight $x_v\ge0$ is a quantity of token $v$, such as its supply, its reserves in pools or a wallet's holdings. Market cap values a token's circulating supply, $\Mstd_i=q_i\,p_i$, and $\Mstd(S)=\sum_{i\in S}\Mstd_i$ for a set $S$ of tokens. Liquidity values the reserves that trade within a fixed band around a token's price in its neighbour pools, commonly $\pm2\%$, and portfolio value values holdings.

In a star market, standard metrics err only through \emph{illiquidity}. A mark is a marginal price, so it overstates what a large sale returns, but this error depends only on a token's own venues and adds up across tokens. We do not study it.

In a graph-structured market, a token's price route passes through upstream nodes, and standard metrics treat each of them as if it were the anchor. Each upstream node $a$ opens three channels into a token's metrics, one each for prices, sales and purchases. A channel distorts the metrics only as far as $a$ departs from the anchor in one of three properties, each measured by an indicator.

\textit{P1: negligible quote-asset risk.}
The quote asset $a$ is far less volatile than the tokens it prices. The USD return of token $i$, $r_i=\dlog p_i$, is its local return $r^\ell_i=\dlog\ell_i$ plus $r_i-r^\ell_i$ imported from its upstream nodes. Its indicator is the \emph{imported volatility share} of each $i\in\down(a)$, the share of the variance of $r_i$ due to the imported part,
\begin{equation}\label{eq:model-share}
s_i\;=\;\frac{\Cov\!\big(r_i-r^\ell_i,\;r_i\big)}{\Var(r_i)} .
\end{equation}

\textit{P2: independent exits.}
The tokens in $\down(a)$ do not compete for the markets through which $a$ exits, so their standalone sale values add up. Its indicator is how far a joint sale falls short of their sum.

\textit{P3: negligible price impact from the tokens it prices.}
Trading the tokens in $\down(a)$ barely moves the price of $a$. Its indicator is the response of $a$'s return to net purchases of these tokens.

A star market has no upstream nodes, so these channels do not exist. A fiat currency, such as the yen for Japanese stocks reported in USD, is an upstream node that nearly satisfies all three properties (\Cref{sec:intro}). Many on-chain quote assets, such as platform hubs and local hubs, fall far short of them, which produces \emph{the pairing biases}.

\section{Three Pairing Biases}\label{sec:biases}

\begin{figure*}[!t]
\centering
\includegraphics[width=\textwidth]{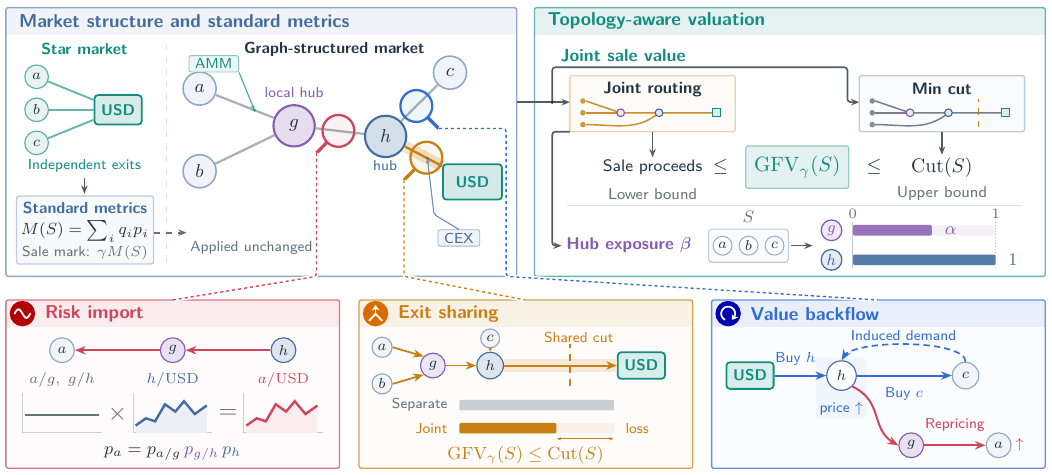}
\caption{Overview of the pairing bias. Standard metrics, built for a star market, are applied unchanged to a graph-structured market in which leaves $a$ and $b$ are priced through the local hub $g$, and $g$ and $c$ through the hub $h$. In risk import, $a$'s USD price $p_a=p_{a/g}\,p_{g/h}\,p_h$ moves with $g$ and $h$ even when its own local rate is flat. In exit sharing, the sales of $a$, $b$ and $c$ cross one shared cut to USD, so a joint sale returns less than separate sales. In value backflow, a USD purchase of $c$ first buys $h$, and the rise in $h$'s price reprices $a$ through $g$. Topology-aware valuation brackets $\mathrm{GFV}_\gamma(S)$ for $S=\{a,b,c\}$ between the proceeds of a routed joint sale and the cut value $\mathrm{Cut}(S)$. The exposure vector reports the share of the proceeds that passes through each hub.}
\label{fig:mechanism}
\end{figure*}

Standard metrics see a market only through supplies and marks, so they cannot tell apart markets with different routes, venues and reserves. \Cref{fig:mechanism} illustrades the mechanisms and the corresponding valuation tools. In a graph-structured market, the upstream nodes on a price route can fail each property of \Cref{sec:anchor}, and each failure produces one pairing bias, \iconI~\RI{} from P1, \iconII~\ES{} from P2 and \iconIII~\VB{} from P3.

\subsection{\texorpdfstring{\iconI\ }{}Bias~I: risk import}\label{sec:risk}
The mark of a token with upstream nodes is a product of the prices along its price route, so the price moves of every upstream node pass into it. For a route $A\to B\to\mathrm{ETH}\to$ USD, $\Mstd_A=q_A\;p_{A/B}\;p_{B/\mathrm{ETH}}\;p_{\mathrm{ETH}}$,
so the local hub $B$ and the designated hub ETH each contribute one factor. Standard metrics report only the product. A hub's volatility, through this product, is imported into the price of every token quoted through it (\Cref{fig:mechanism}).

\textit{Decomposition.}
Let token $i=v_0$ have upstream nodes $v_1,\dots,v_n$ in route order, and take returns over a period in which the price forest does not change. Taking logarithms of the mark gives
\[
\dlog\Mstd_i\;=\;\dlog q_i+\sum_{k=0}^{n}r^\ell_{v_k}.
\]
By the bilinearity of covariance, $\Var(r_i)=\sum_{k}\Cov(r^\ell_{v_k},r_i)$, so the level shares $\Cov(r^\ell_{v_k},r_i)/\Var(r_i)$ sum to one, and $s_i$ is the sum of the upstream shares. When a price feed averages several routes with weights $w_\rho$, a 1\% move in the local rate of a node $v$, with the other local rates fixed, moves $p_i$ by $\beta_i(v)$\%, where the \emph{exposure} $\beta_i(v)=\sum_{\rho\ni v}w_\rho\in[0,1]$ is the total weight of the routes through $v$. Arbitrage between the routes then moves the local rates themselves.

\textit{What standard metrics misread.}
Reading the product alone misreads a token in three ways. If a hub's USD price doubles while everything below it stays fixed, the market cap of every token it prices doubles, so rankings across hubs move with relative hub prices. If a token's own local rate stays fixed, all of its market-cap move comes from upstream, yet a standard metric reports it as a move of the token. Holdings $x$ carry the exposure $\sum_i x_i\,p_i\,\beta_i(v)$ to each node $v$, so a portfolio spread over the tokens of one hub is concentrated on that hub. For example, VIRTUAL agent tokens are priced through VIRTUAL and ETH and vary along with them, and when ETH or BNB falls by 10\% or more, the USD prices of tokens quoted in it fall almost one for one while their local rates barely move (\Cref{fig:risk-events}). Two corrections follow. A metric can report market cap in units of an upstream node $v$, $\Mhub{v}_i=q_i\,p_{i/v}$, where $p_{i/v}$ is the product of the local rates from $i$ up to, but excluding, $v$. Since $\Mstd_i=\Mhub{v}_i\,p_v$, this removes $v$ and every node above it, as constant-currency reporting does for a local currency. Or it can report its exposures to each upstream node, as \Cref{sec:gfv} does.

\subsection{\texorpdfstring{\iconII\ }{}Bias~II: exit sharing}\label{sec:exit}
A token without a deep market against USD exits along its route. For example, a memecoin quoted in a hub is first sold for the hub, and the hub is then sold for USD, so the memecoin borrows the hub's channel to USD. The hub's own holders and every other token priced through the hub share the same channel. When they sell together, they compete for it, and each realizes less than it would alone. We call this \emph{exit sharing}. Even a token sold alone is limited by the thinnest leg of the channel it borrows, and the shared channel can sit at any upstream node (\Cref{fig:mechanism}).

A \emph{liquidation plan} for a set $S$ sells at most a share $\gamma\in(0,1]$ of the supply of every token in $S$ through a sequence of trades, and its proceeds are the sink set (USD and stablecoins). The \emph{joint realizable value} $J_\gamma(S)$ is the most a plan can return. We compare three tiers, the standard value of the amount sold $\Tone=\gamma\,\Mstd(S)$, the sum of the standalone sales $\Ttwo=\sum_{i\in S}J_\gamma(\{i\})$, and the joint sale $\Tthree=J_\gamma(S)$. $\Tone-\Ttwo$ is what the tokens lose even when each is sold alone. $\Ttwo-\Tthree$ is what they lose to each other by shared exits. 
The joint realizable value is hard to compute, but a pool can pay out at most its reserves and a book can absorb at most its bids, so the value that can flow from $S$ to the sink set is limited. Give each pool edge $u\to v$ the capacity $p_v\,R_P(v)$, the reserve it pays out valued at its mark, and each book edge the depth $D_b$ of its bids. The \emph{cut value} $\Cut(S)$ is the maximum flow along these edges from the tokens in $S$, each supplying at most its market cap, to the sink set. By max-flow min-cut, it equals the capacity of the narrowest set of edges that separates $S$ from the sinks, the \emph{exit cut}. 

\textit{Cut bound.}
In a \emph{no-arbitrage state}, where pool prices agree with the marks and no bid exceeds them, three facts follow. (i) The cut value bounds any sale, $J_\gamma(S)\le\min\{\gamma\,\Mstd(S),\Cut(S)\}$. (ii) Unlike market cap, the cut value does not add up, $\Cut(S\cup S')\le\Cut(S)+\Cut(S')$, since $S$ and $S'$ may compete for the same bottleneck. Non-additivity under illiquidity is known~\cite{Acerbi2008LiquidityRisk}, but here it couples distinct tokens. (iii) Leaves that exit only through their hub add market cap but no cut value, since any cut that separates the hub from the sinks also separates them, so the hub and its leaves share one cut.

\textit{Illiquidity versus pairing.}
The gap $\Tone-\Ttwo$ has two sources: the illiquidity of the tokens' own pools and the illiquidity of the upstream legs they borrow. The former also exists in star markets while the latter is a pairing bias. To separate them, we split each sale into legs. A leg that sells $x$ units of $a$ for $y$ units of $b$ loses $p_ax-p_by$, and the losses of a completed sale sum to the standard value sold minus the proceeds. In the standalone sale of $i$, the legs that sell $i$ form its \emph{own-market loss}, and the legs that sell an upstream node $v$ form the \emph{upstream loss} at $v$. The \emph{pairing-specific loss} is the upstream loss plus the shared-exit gap $\Ttwo-\Tthree$. It is zero in a star market.

\textit{What standard metrics misread.}
Standard metrics treat each token's exit as its own, so summed market cap, liquidity and portfolio value overstate what a set can be sold for, and liquidity counts one hub exit once per leaf. CoinGecko even sums market cap over whole ecosystems~\cite{CoinGeckoVirtualsCategory}, whose tokens share one hub exit. For example, ten Ethereum tokens priced through ELEVATE, whose own price rests on \usd{17.7}k of ETH, carry \usd{1.91}bn of float market cap, yet with ELEVATE their cut value is \usd{99.7}k.

\subsection{\texorpdfstring{\iconIII\ }{}Bias~III: value backflow}\label{sec:backflow}
To buy a leaf quoted by a hub, one must first buy the hub with USD and then the leaf with the hub. So demand for the leaf raises the hub's price. Seeding leaf-hub pools also locks hub tokens. Value created downstream thus flows back into the hub, and through it into other downstream tokens. We call this \emph{value backflow} (\Cref{fig:mechanism}).

\textit{Hub response and amplification.}
Take the price route $T_n\to\dots\to T_1\to T_0=$ USD of a token $T_n$, and suppose each $T_k$ on it is also priced along the rest of this route, so $T_1$ is a hub when $n\ge2$. Pool $j$ trades $T_j$ against $T_{j-1}$, and its exchange function gives its \emph{price impact} $\eta_j$, the relative rise of its local price per dollar spent in it. For example, a constant-product pool has $\eta_j=2/L_j$, with $L_j$ the USD value of its $T_{j-1}$ reserve. Buying $X$ dollars of $T_n$ along the route first buys $T_1$ in pool~1. Each pool $j$ then receives about $X$ dollars and its local price rises by $\eta_jX$, so the mark of $T_k$ rises by about $X\sum_{j\le k}\eta_j$. The combined market cap of all tokens on the route, $T_1$ to $T_n$, therefore rises by about $A_0X$ (Eq.~\ref{eq:biases-amp}). Only the term $j=n$, $\eta_n\Mstd_n$, in which the purchased token reprices in its own pool, would arise in a star market. The terms with $j<n$, in which each hub's pool reprices and lifts every token below it, are the backflow.
\begin{equation}\label{eq:biases-amp}
A_0\;=\;\sum_{j=1}^{n}\eta_j\sum_{k\ge j}\Mstd_k.
\end{equation}

\textit{What standard metrics misread.}
Standard metrics book a hub's market cap as its own worth, although part of it rests on leaves being bought through the hub. One dollar spent on a leaf thus adds about $A_0$ dollars of market cap across the route. This demand is real, as memecoin traders on pump.fun need SOL to trade. Foreign buyers of Japanese stocks likewise need yen, but the yen market is so deep that they barely move it, whereas leaf flows can move a hub's price far. The problem is that this part of a hub's value depends on leaf flows, which standard metrics do not show.

\subsection{Exploiting the biases}\label{sec:strategic}

Issuers and platforms choose their topology and can exploit the three biases to inflate their tokens' standard metrics, which we call \textbf{topology manipulation}. Through Bias~I, a token can borrow a hub's volatility. A token quoted in USD keeps its price without trades, but a token quoted in ETH moves accordingly, so in a bull market it rises with ETH without new demand and looks popular.

Through Bias~II, an issuer can inflate market cap with almost no capital. It mints tokens $A$ and $B$, pools them deeply against each other, and links $B$ to a stablecoin through a pool of $\varepsilon$ dollars. The market cap of $A$ can then be arbitrarily large, yet $A$ and $B$ together sell for at most $\varepsilon$. Many tokens can be priced through the same $B$. Terra had this shape, with the conversion between LUNA and UST as the deep pool~\cite{Liu2026AnatomyCrash,Briola2023TerraLuna}. On 1 May 2022, UST and LUNA had a combined market cap of \usd{46.0}bn, but their Ethereum exit pools held only 1.5\% of that (CEX not counted).
With Bias~III, a platform can make its own token the hub of other tokens, so that their demand becomes demand for its token. Virtuals and Zora do so by placing their own tokens on the routes of their launches.

\section{Topology-Aware Valuation}\label{sec:metrics}\label{sec:gfv}
A topology-aware metric keeps the inputs of the standard metrics, venues and marks, but must meet two requirements. (i) It must be \emph{exit-aware}, valuing a sale by the set's joint exits to USD (Bias~II). In a star market this reduces to the usual liquidity adjustment~\cite{Acerbi2008LiquidityRisk}. (ii) It must be \emph{hub-aware}, reporting exposure to each upstream node without folding it into one number (Bias~I). Value backflow can only be disclosed. We propose GFV for the first requirement and an exposure vector for the second (\Cref{fig:mechanism}).

\textit{Generalized flow valuation.}
GFV writes the joint sale of \Cref{sec:exit} as a flow through the venues. Let $x^{uv}_P\ge0$ be the amount of $u$ sold into pool $P$ for $v$, which returns $\Psi^{uv}_P(x^{uv}_P;R_P)$ of $v$. A \ac{CEX} book $b$ enters as a pool of $v(b)$ and $\mathsf{usd}$ whose exchange function fills its bids from the best down, up to $D_b$. For a set $S$ of non-sink tokens and a liquidation parameter $\gamma\in(0,1]$,
\begin{equation}\label{eq:metrics-gfv}
\begin{aligned}
\GFV_\gamma(S)=\max_{x\ge0}\ &\sum_{u\in\mathcal U}\sum_{P,v}\Psi^{vu}_P(x^{vu}_P;R_P)\\
\text{s.t.}\ &\sum_{P,v}x^{iv}_P\le\gamma q_i+\sum_{P,v}\Psi^{vi}_P(x^{vi}_P;R_P),\ \ i\in S,\\
&\sum_{P,v}x^{iv}_P\le\sum_{P,v}\Psi^{vi}_P(x^{vi}_P;R_P),\ \ i\notin S\cup\mathcal U,\\
&x^{uv}_P=0,\ \ u\in\mathcal U.
\end{aligned}
\end{equation}
Tokens in $S$ sell at most a share $\gamma$ of their supply plus what they receive, other tokens only pass through, and sinks are never sold. Every $\Psi$ is concave, so this is a convex optimal routing problem.

\textit{Properties.}
In a no-arbitrage state, a sale returns at most the marked value it sells. Every dollar it returns also crosses the exit cut, so $\GFV_\gamma(S)\le\min\{\gamma\Mstd(S),\Cut(S)\}$. Shrinking all flows by a factor $t\le1$ keeps them feasible and, since $\Psi$ is concave, returns at least $t$ times as much, so $\GFV_\gamma(S)/\gamma$ rises as $\gamma$ falls. As $\gamma\to0$ each token sells along its best route at marginal rates, so $\GFV_\gamma(S)/\gamma$ tends to $\Mstd(S)$. Market cap is thus the zero-stress limit of GFV, pricing every unit at the marginal price. Like the cut value, GFV is subadditive, as a joint sale returns no more than the separate sales.

\textit{Manipulation.}
Every dollar a sale of $S$ returns must leave through the venues that trade its tokens, so GFV never exceeds what these venues hold of other tokens. Pools, trades and mints inside $S$ do not change these holdings, so they cannot lift GFV above them, although they can inflate market cap without bound. They can still raise GFV up to this cap, as when a pool links a token without exits to an unused exit, since that liquidity is real. 

\textit{Exposure.} Next to GFV, we report how its proceeds depend on each upstream node. Decompose the sale into paths $\rho$, each returning $g_\rho$ dollars, and let $\beta^\gamma_S(v)=\sum_{\rho\ni v}g_\rho/\GFV_\gamma(S)$ be the share of the proceeds that passes through $v$. As $\gamma\to0$ it tends to the market-cap-weighted exposure of \Cref{sec:risk}, but a token with large market cap and thin liquidity carries little weight. 

\begin{table*}[htb]
\footnotesize
\caption{Standard and topology-aware metrics on five hubs and their leaves. (I) \emph{Hub share}: share of daily changes in the cohort's total / each token's market cap (value-weighted). (II) $\Tone$: market value sold; $\Ttwo$, $\Tthree$: proceeds of separate and joint sales ($\Tthree$ estimates GFV). \emph{Added gap}: extra loss, over $\Ttwo$, when the hub's holders also sell $\gamma$. \emph{Cut ratio}: hub in leaf pools over the hub's uncapped exit capacity to USD. (III) \emph{Hub in leaf pools}: share of the hub's \acs{AMM} inventory / supply. \emph{Hub price change}: price rise after buying $\gamma$ of the hub in leaf pools at once.}
\label{tab:compare}
\begin{tabular*}{\textwidth}{@{\extracolsep{\fill}}p{3.6cm}lllll@{}}
\toprule
 & ETH leaves & ETH leaves & VIRTUAL leaves & ZORA leaves & BNB leaves\\
 & (Ethereum) & (Base) & (Base) & (Base) & (BSC)\\
\midrule
\multicolumn{6}{@{}l@{}}{\textit{Risk import} \bI: leaves tracked daily from Jan.\ 2025 to Sep.\ 2026, market cap at the snapshot (3,041 / 965 / 277 / 22 / 7,161 tokens)}\\
Market cap, USD / hub units & \usd{123.6}bn / 46.0M ETH & \usd{42.0}bn / 15.6M ETH & \usd{555.0}M / 689M VIRTUAL & \usd{1.52}M / 191M ZORA & \usd{82.4}bn / 107M BNB\\
Hub share, total / per token & 77.1\% / 59.9\% & 51.2\% / 40.6\% & 61.9\% / 45.2\% & 41.6\% / 34.8\% & 75.8\% / 58.0\%\\
\midrule
\multicolumn{6}{@{}l@{}}{\textit{Exit sharing} \bII: all screened leaves at the snapshot, data at $\gamma=1\%$ / $10\%$ (3,317 / 1,408 / 273 / 24 / 3,602 tokens)}\\
Market cap, sum (standard) & \usd{167.1}bn & \usd{43.9}bn & \usd{271.0}M & \usd{1.54}M & \usd{79.1}bn\\
GFV / market cap, $\Tthree/\Tone$ & 16.5\% / 2.37\% & 5.89\% / 1.21\% & 50.1\% / 15.7\% & 94.6\% / 72.5\% & 8.18\% / 1.48\%\\
Shared-exit gap, $(\Ttwo-\Tthree)/\Ttwo$ & 0.34\% / 1.43\% & 10.9\% / 18.0\% & 0.53\% / 5.81\% & 0.44\% / 1.13\% & 10.1\% / 28.0\%\\
Added gap from hub's sale & 26.1\% / 41.3\% & 0.04\% / 4.86\% & 15.2\% / 82.8\% & 11.9\% / 82.1\% & 44.3\% / 45.8\%\\
Cut ratio & 0.53$\times$ & 0.16$\times$ & 1.07$\times$ & 0.20$\times$ & 1.45$\times$\\
\midrule
\multicolumn{6}{@{}l@{}}{\textit{Value backflow} \bIII: hubs at the snapshot, data at $\gamma=0.1\%$ / $1\%$}\\
Hub in leaf pools, \acs{AMM} / supply & 52.9\% / 0.17\% & 38.6\% / 0.031\% & 77.8\% / 2.34\% & 82.9\% / 0.90\% & 82.4\% / 0.55\%\\
Hub price change & 0.23 / 3.7\,\acsp{bp} & 0.07 / 0.61\,\acsp{bp} & 3.7 / 16.6\,\acsp{bp} & 1.9 / 13.8\,\acsp{bp} & 3.6 / 51.5\,\acsp{bp}\\
\bottomrule
\end{tabular*}

\end{table*}

\section{Measurement}\label{sec:measurement}

\subsection{Data and methodology}\label{sec:data}
We read the pools of the whitelisted \ac{DEX} factories on Ethereum, Base and BSC on 2026-10-01 and keep those with at least \$1k of liquidity (43,236, 12,780 and 53,200 pools). Each token is priced in its deepest pool, and its supply $q_i$ is a holder-adjusted float, total supply minus identified burn, bridge, treasury and platform balances. A token's pool backing is the USD value of the other tokens held in its pools, each capped by the liquidity behind its own price. Time series end at this snapshot and come from on-chain events, also on Robinhood Chain, Binance hourly closes, \ac{CMC} daily listings~\cite{coinmarketcap} and stock prices.

\textit{Estimating GFV.}
We sell on \ac{DEX} pools with their exact formulas or a fork of the deployed contracts, and on the order books of 12 \acp{CEX} at the snapshot. Each sale of a share $\gamma\in\{0.1\%,1\%,5\%,10\%\}$ is split into chunks, each sold along the best path of at most four swaps. These sales are feasible, so their proceeds are lower bounds on GFV. Making the routing optimal is outside our scope~\cite{Angeris2022OptimalMakers,Angeris2022ConstantOptimization}.

\subsection{The graph topology of on-chain markets}\label{sec:prevalence}

\begin{table}[htb]
\footnotesize
\setlength{\tabcolsep}{1.6pt}
\caption{Tokens by the first upstream node of their price route, in \% of tokens, of float market cap and of pool backing, excluding Volatile hub (ETH, BNB, BTC and their staked and wrapped versions) and stablecoins. Platform hub: launch-platform and \acs{DEX} tokens such as VIRTUAL, ZORA and AERO. $^\dagger$Omitted, as market cap on BSC is distorted by tokens with extreme marks.}
\label{tab:prev-routes}
\begin{tabular}{@{}lrrrrrrrrr@{}}
\toprule
& \multicolumn{3}{c}{Ethereum} & \multicolumn{3}{c}{Base} & \multicolumn{3}{c}{BSC}\\
\cmidrule(lr){2-4}\cmidrule(lr){5-7}\cmidrule(lr){8-10}
& Tok. & Float & Back. & Tok. & Float & Back. & Tok. & Float$^\dagger$ & Back.\\
\midrule
USD & 7.0 & 23.7 & 21.2 & 11.7 & 40.5 & 33.6 & 21.7 & -- & 38.6\\
Volatile hub & 91.2 & 74.7 & 75.1 & 62.2 & 57.5 & 55.4 & 72.4 & -- & 55.0\\
Platform hub & 0.0 & 0.0 & 0.0 & 20.2 & 0.4 & 6.6 & 0.0 & -- & 0.0\\
Tokenized stock & 0.1 & 0.0 & 0.0 & 0.0 & 0.0 & 0.0 & 2.6 & -- & 0.4\\
Local hub & 1.7 & 1.6 & 3.7 & 6.0 & 1.6 & 4.5 & 3.3 & -- & 6.0\\
\midrule
Any non-USD node & 93.0 & 76.3 & 78.8 & 88.3 & 59.5 & 66.4 & 78.3 & -- & 61.4\\
$\geq$2 paired tokens & 11.2 & 56.3 & 55.5 & 14.1 & 11.8 & 65.0 & 9.0 & -- & 51.7\\
Tokens & \multicolumn{3}{c}{8,287} & \multicolumn{3}{c}{3,084} & \multicolumn{3}{c}{13,123}\\
\bottomrule
\end{tabular}
\end{table}

We follow each token's price route to USD and classify it by its first upstream node (\Cref{tab:prev-routes}), excluding ETH, BTC, BNB, their wrappers and stablecoins. Graph structure is the norm, as tokens with pools against at least two other tokens hold 51.7--65.0\% of the pool backing. Pairing with tokens other than stablecoins is also prevalent, as 78.3--93.0\% of tokens and 59.5--76.3\% of value are priced through at least one non-USD node. Platform and local hubs price many tokens but carry little value.

USD stablecoins quote a rising share of new pools, 23.5--31.8\% of new leaf pools in 2026, but new hub layers keep appearing. On Base, 36.1\% are quoted in other tokens, mostly Zora content coins quoted in their creator's coin, which is quoted in ZORA, and on BSC tokenized stocks quote 13.5\%.

\subsection{Standard vs. topology-aware metrics}\label{sec:compare}
\Cref{tab:compare} compares standard and topology-aware metrics on five hubs and their leaves, namely the native hubs ETH (on Ethereum and Base) and BNB (on BSC) and the platform hubs VIRTUAL and ZORA (on Base). Its three blocks measure how far each hub departs from properties P1, P2 and P3 of a star-market anchor (\Cref{sec:anchor}).

\noindent\textit{Bias~I\@.} The hub's price accounts for 41.6--77.1\% of the daily changes in each cohort's market cap, and for 34.8--59.9\% per token. Reporting market cap in hub units removes this imported volatility. For VIRTUAL leaves, the total share of 61.9\% splits into 38.4 points from VIRTUAL's rate against ETH and 23.5 from ETH's USD price.

\noindent\textit{Bias~II\@.} At $\gamma=1\%$ the joint sale, by our GFV estimation, realizes only 5.89--16.5\% of the market cap sold for the ETH and BNB cohorts. Much of this shortfall is own-market slippage, which a liquidity adjustment also captures in star markets, but pairing adds a loss that no per-token adjustment sees. This loss is small when only the leaves of deep hubs are sold, but grows with stress. When the hub is also sold, shared exits cost the leaves a further 11.9--44.3\% of their standalone proceeds, as the hub's own sale takes most of its exit depth, except for ETH on Base. At $\gamma=10\%$ the leaves of BNB and of ETH on Base lose 28.0\% and 18.0\% without the hub's sale.

\noindent\textit{Bias~III\@.} We have no topology-aware metric for this bias. Leaf pools hold about 80\% of the \ac{AMM} inventory of VIRTUAL, ZORA and BNB, against about half for ETH, so the same relative inflow into leaf pools moves these hubs 4 to 84 times as much as ETH.

\subsection{\texorpdfstring{\iconI\ }{}Bias~I: risk import}\label{sec:mag-risk}
\begin{figure}[htb]
 \centering
 \includegraphics[width=\linewidth]{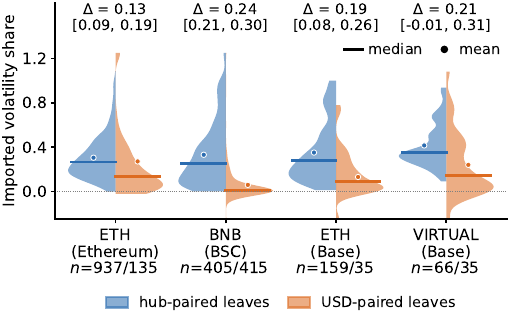}
 \caption{Imported volatility share per token-year, from 4-hour USD returns, 2023--2026. USD-paired leaves are measured against the same hub. $\Delta$: median difference with its 95\% bootstrap interval; $n$: token-years, hub-paired / USD-paired.}
 \label{fig:risk-share}
\end{figure}

\textit{Imported volatility share.}
\Cref{fig:risk-share} compares the share $s_i$ of \eqref{eq:model-share}, estimated from 4-hour USD returns, for leaves quoted in a hub and for USD-paired leaves measured against the same hub. For every hub, hub-paired leaves import more of the hub's volatility than USD-paired leaves, with a median share of 0.25 to 0.35, in every year from 2023 to 2026 (\Cref{app:risk}). The groups are not matched, and the common market factor of crypto assets~\cite{Liu2022CommonFactors} also moves some USD-paired tokens with the hub.

\begin{figure}[htb]
 \centering
 \includegraphics[width=\linewidth]{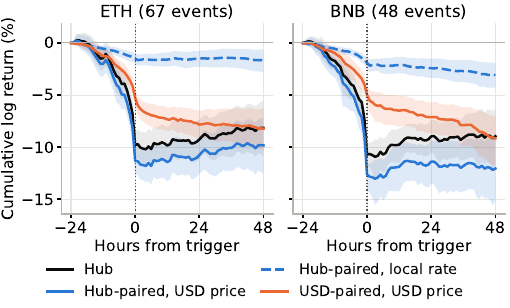}
 \caption{Mean cumulative log return around ETH and BNB drawdowns of at least 10\%, 2021--2026, for Ethereum and BSC leaves. Tokens are averaged within each event, then over events; bands are 95\% bootstrap intervals.}
 \label{fig:risk-events}
\end{figure}

\textit{Hub drawdowns.} We take every time ETH or BNB price drops greatly (closed at least 10\% below its 1-day high). Shown in \Cref{fig:risk-events}, at the trigger, hub-paired leaves have fallen 6--8 log points more than USD-paired leaves, though their local rates against the hub have fallen less than 2 points. Most of their fall comes from the hub.

\textit{Market cap and rankings.}
At the snapshot, 48.6\% of the market cap of ETH leaves on Ethereum sat in pools without a price-moving trade within 24 hours, yet their USD prices kept moving with ETH. These also move rankings, as the ETH price explains 26.8\% of the 30-day rank changes of ETH leaves in the \ac{CMC} top 5,000.

\subsection{\texorpdfstring{\iconII\ }{}Bias~II: exit sharing}\label{sec:mag-exit}

\begin{figure*}[htb]
 \centering
 \includegraphics[width=\linewidth]{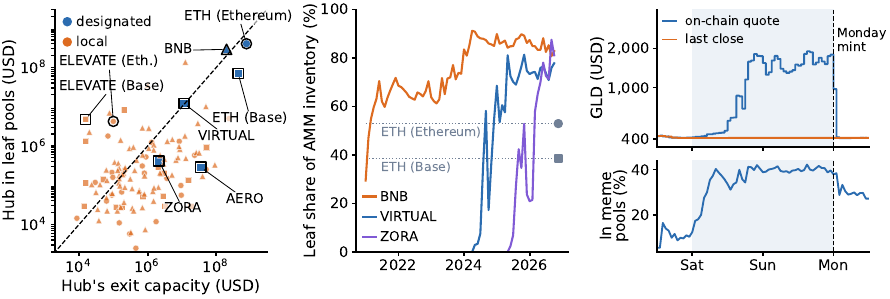}
 \caption{Exit sharing and value backflow. Left: for each hub, the hub held in its leaves' pools against the hub's uncapped exit capacity to USD. Above the dashed line, selling the leaves alone brings more of the hub than its exit can convert. Circles, squares and triangles mark Ethereum, Base and BSC. Middle: month-end share of each hub's \acs{AMM} inventory held in leaf pools, with ETH at the snapshot for reference. Right: the GLD stock token on Robinhood Chain over the weekend of 29 August 2026, with its on-chain quote against the last close and the GLD held in memecoin pools.}
 \label{fig:bias2}\label{fig:bias3}
\end{figure*}

\textit{Hub ecosystems.}
 Beyond the five hubs of \Cref{tab:compare}, \Cref{fig:bias2} (left) compares, for every hub, the hub held in its leaves' pools with the hub's own exit capacity to USD, whose ratio is the cut ratio of \Cref{tab:compare}. Hubs below the line, such as ETH, AERO and ZORA, have room to spare. For VIRTUAL and BNB, leaf sales alone can fill the hub's exit. Local hubs with at least three leaves or \usd{1}M of downstream float market cap have far thinner exits, and 36 of the other 117 hubs lie above the line (\Cref{tab:app-hubs}), 9 of them more than 10 times, such as ELEVATE of \Cref{sec:exit} at 42 times, whose tokens lose 19.1\% of $\Ttwo$ to the shared exit when sold together at $\gamma=1\%$.

\textit{What pairing adds to thin liquidity.}
Thin liquidity inflates market cap~\cite{Mongardini2025MemeManipulations}, and would do so in a star market too. Pairing adds two things. A token's pool against its hub can be deep while the hub itself is thin, so its own pools hide how little it can be sold for (topology manipulation in \Cref{sec:strategic}). And many tokens can share one thin hub, so their summed market cap rests on one exit. Tokens priced through local hubs whose leaves carry more than 100 times their cut value hold \usd{1.2}--10.0bn of float market cap per chain, 0.5--2.9\% of the total. On BSC, another 48 tokens priced through tokens whose contracts block sales carry \usd{23.5}bn on \usd{157}k of cut value.

\subsection{\texorpdfstring{\iconIII\ }{}Bias~III: value backflow}\label{sec:mag-backflow}

\textit{Downstream liquidity.}
Leaf pools hold most of the \ac{AMM} inventory of VIRTUAL, ZORA and BNB (\Cref{tab:compare}), a share that has risen over time (\Cref{fig:bias3}, middle), and on BSC eight tokenized stocks each price 134 to 1,105 memecoins. The inflows simulated in \Cref{tab:compare} are of a size seen daily, as the median absolute daily net flow into leaf pools in July--September 2026 was 0.09--1.08\% of the hub held there.

\textit{Hub price vs. leaf inflows.}
Each hour, we regress the hub's return on the net amount of the hub that buyers pay into its leaf pools, as a share of supply, with Newey--West standard errors. Before each hub's first Binance perpetual listing, a net inflow of 0.01\% of supply comes with a 0.68\% rise in VIRTUAL's price and a 0.75\% rise in ZORA's (both $p<0.01$, significant; \Cref{tab:backflow-flows}). After the listing, the positive relation disappears, consistent with \acp{CEX} taking over price discovery, as for BNB, which trades on \acp{CEX} throughout.

\textit{Tokenized stocks.}
Tokenized stocks are minted and redeemed on weekdays only, but memecoins quoted in them trade all weekend, so buying these memecoins can raise the price of the stock token they are quoted in. Over the last weekend of August 2026, buyers of GLD-quoted memecoins drew GLD into their pools, from 10\% to 42\% of its fixed supply, and its quote rose to an hourly median of \usd{1,382} against a Friday close of \usd{409} until Monday (\Cref{fig:bias3}, right). Across 105 stock tokens the evidence is weaker, as a share in memecoin pools 10 points higher comes with 0.8\% more deviation from the last close on weekends than on weekdays ($p=0.10$).

\section{Discussion}\label{sec:discussion}

\textit{Recommendations.}
The biases arise where metrics are produced, so the fixes belong to the platforms and websites that publish them. For Bias~I, aggregators and wallets can show a token's market cap in units of its hub next to its USD market cap, together with the exposure vector $\beta$. For Bias~II, listing criteria and lending protocols can value a token, or a set of tokens that share an exit, by its \ac{GFV} at a given $\gamma$. For Bias~III, which has no corrective metric, platforms can disclose the share of a hub's \ac{AMM} inventory held in its leaves' pools. Since issuers choose their pairings, the metrics that platforms report can in turn shape these choices. Reporting realizable value limits topology manipulation, as pools and trades inside a set of tokens cannot lift its \ac{GFV} above what its exits hold (\Cref{sec:gfv}).

\textit{Limitations.} Three limitations qualify our results.
\begin{itemize}[leftmargin=*,itemsep=1pt,topsep=2pt]
 \item \emph{Snapshot data.} The structural measurements, from prevalence to sale proceeds and cut values, are read at one snapshot, 2026-10-01 00:00~UTC, so a sale cannot see market makers quoting or cancelling, or \ac{AMM} liquidity moving, while it proceeds. \ac{CEX} books enter as the visible spot depth of the exchanges we cover, and the \ac{DEX} venues we read hold 90.8--96.7\% of the \ac{DEX} \ac{TVL} that DefiLlama~\cite{defillama} reports on each chain.
 \item \emph{Self-selection.} Issuers choose their pairings, so differences between hub-paired and USD-paired tokens may reflect which tokens choose a hub, and we report them as associations.
 \item \emph{No causal estimate.} Leaf inflows and hub returns rise together (\Cref{sec:mag-backflow}) for two possible reasons. Demand for leaves may lift the hub, or attention to a project may draw buyers to both. Our data cannot separate them. That the relation disappears once the hub trades on a \ac{CEX} is consistent with the pairing channel, though the listing also comes as each hub matures. The weekend evidence on Robinhood Chain is significant only at the 10\% level.
\end{itemize}

These limitations point to next steps. Realizable value can be measured as liquidity responds to a sale, and natural experiments can identify how far demand for leaves moves their hubs.

\section{Related Work}\label{sec:related}
\textit{Vehicle currencies and international finance.} Foreign exchange is a graph-structured market whose hub is a vehicle currency, the dollar~\cite{Krugman1980Vehicle,Devereux2013Vehicle}. This literature explains why a vehicle emerges and persists~\cite{Matsuyama1993TowardCurrency,Rey2001InternationalExchange,Devereux2013Vehicle,He2016NetworkCurrency} and which currency invoices trade~\cite{Goldberg2008Vehicle}. Related work studies how dollar moves pass into trade prices~\cite{Gopinath2020DominantParadigm}, how shared markets spread fire sales and liquidity spirals~\cite{Cifuentes2005LiquidityContagion,Brunnermeier2009FundingLiquidity}, and how convenience demand raises the dollar's value~\cite{Gopinath2021DominantCurrency,Jiang2021ForeignSafe}. These studies ask how the hub affects prices and real outcomes. We ask instead how metrics that treat a hub as the unit of account misread the assets priced through it, and trace all three distortions to the market topology.

\textit{Financial networks.} Systemic-risk models spread losses between institutions through claims~\cite{Eisenberg2001SystemicRisk,Acemoglu2015SystemicNetworks,Elliott2014FinancialContagion,Rogers2013FailureRescue} or a commonly held asset~\cite{Cifuentes2005LiquidityContagion}, whereas our graph couples distinct assets through the pools on their routes to USD.

\textit{Liquidity-adjusted valuation and optimal liquidation.} Under illiquidity, portfolio value is a nonlinear function of the holdings and of the investor's liquidity policy~\cite{Acerbi2008LiquidityRisk}, and optimal-execution models minimize price-impact costs~\cite{Bertsimas1998OptimalCosts} or trade them against timing risk~\cite{Almgren2001OptimalTransactions}, with a cross-impact matrix through which trading one asset moves another's price. Non-additivity is therefore not new, but there the coupling is a parameter. In exit sharing it comes from the graph, as assets reach USD through shared upstream nodes, and the cut value bounds their joint proceeds by a minimum USD cut.

\textit{\acsp{AMM} and \acs{DeFi} measurement.} \ac{CFMM} theory finds the best execution of a trade across pools, including the sale of a basket of tokens, by convex optimization~\cite{Angeris2022ConstantOptimization,Angeris2022OptimalMakers}. We do not study how to route, but take the proceeds of the best sale as what holdings can realize and ask how far standard metrics depart from it. Most Uniswap tokens trade directly against WETH in a core--periphery network~\cite{Lehar2025UniswapAMM,Yan2025UniswapNetwork}, the graph on which the biases act. Unlike Lehar and Parlour~\cite{Lehar2025UniswapAMM}, who count tokens, we weight by value and follow routes of every depth. \ac{TVL} counts nested claims more than once~\cite{Luo2026PiercingTVL}, and market cap does the same with wrapped and staked claims, as equity market cap does with cross-holdings~\cite{French1991WereHigh}. Double counting is a different problem, and the pairing biases remain even when every claim is counted once. According to the closest work, 88.1\% of memecoins with a reported cap above \$1B have liquidity below 0.1\% of it, as small trades in thin pools inflate their prices~\cite{Mongardini2025MemeManipulations}. Other measurement work documents wash trading~\cite{Victor2021WashTrading,Cong2023WashTrading}, manipulation and sniper bots~\cite{Szwajcok2026MemeFactories,Cernera2023SniperBots} and scam tokens~\cite{Xia2021ScamTokens}. These distort the prices that standard metrics read, whereas the pairing bias arises even when every price is honest. This also happens in star markets, so it is not a pairing bias, and \Cref{sec:mag-exit} separates what pairing adds to it. Models of platform tokens explain how users' demand gives a token its value~\cite{Cong2021Tokenomics,Sockin2023CryptoModel}. We study where this demand lands on the market graph, as buying a leaf also lifts every hub on its route.

\section{Conclusion}\label{sec:conclusion}
Cryptocurrency web platforms inherit the standard metrics of traditional finance, which ignore what the topology of a graph-structured market means for pricing and valuation. Applied to on-chain markets, these metrics introduce the pairing bias in three forms, risk import, exit sharing and value backflow. Issuers, who choose the pairings, can exploit it to inflate their tokens' metrics. We formalized the pairing bias, measured it on three chains, and proposed topology-aware metrics that report hub exposure and realizable value. The pairing bias arises in any graph-structured market, especially in \acp{AMM}. As more assets, including tokenized stocks, trade on chain and join this topology, such metrics will let traders and platforms see the risk and the realizable value behind the numbers they rely on.

\bibliographystyle{ACM-Reference-Format}
\bibliography{reference}

\appendix
\crefalias{section}{appendix}\crefalias{subsection}{subappendix}\crefalias{subsubsection}{subsubappendix}

\section{Additional Results}\label{app:results}

\subsection{Risk import}\label{app:risk}

\begin{figure*}[t]
\centering
\includegraphics[width=0.75\linewidth]{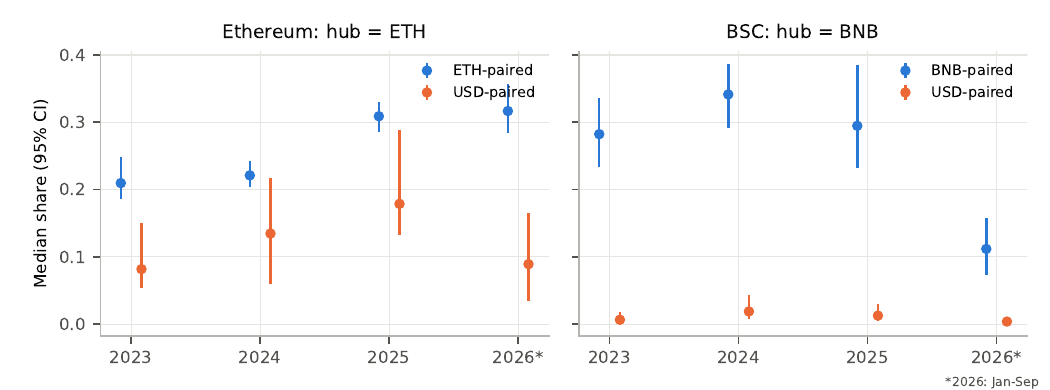}
\caption{The volatility gap holds in every year: median share of a token's USD return variance that comes from its hub, hub-paired and USD-paired tokens, with 95\% \acsp{CI}. 2026 covers January to September.}
\label{fig:app-risk-years}
\end{figure*}

\textit{Imported volatility share.} The gap of \Cref{sec:mag-risk} holds in every year (\Cref{fig:app-risk-years}). From 2023 to 2026, the median hub-paired token imports more of its return variance from the hub than USD-paired tokens share with it, on Ethereum and on BSC. The share of ETH-paired tokens rises from 0.21 in 2023 to 0.32 in 2026, and that of BNB-paired tokens falls to 0.11 in 2026, still far above the 0.004 of USD-paired tokens. The gap remains with daily returns and among tokens whose prices update most often, at 0.10 on Ethereum, where the interval for frequently updated prices includes zero, and 0.21 on BSC (\Cref{tab:app-risk-share}).

\begin{table}[H]
\small\setlength{\tabcolsep}{3.5pt}
\caption{Hub-paired tokens import more of their volatility from the hub than USD-paired tokens, pooled 2023--2026: median shares and their difference with 95\% \acs{CI}.}
\label{tab:app-risk-share}
\begin{tabular}{@{}lccc@{}}
\toprule
Variant & Hub-paired & USD-paired & Difference \\
\midrule
\multicolumn{4}{@{}l}{\emph{Ethereum (ETH)}} \\
4-hour returns & .266 & .132 & .133 [.086, .185] \\
Daily returns & .293 & .193 & .101 [.038, .137] \\
Frequently updated prices & .269 & .165 & .104 [$-$.004, .155] \\
\multicolumn{4}{@{}l}{\emph{BSC (BNB)}} \\
4-hour returns & .252 & .008 & .244 [.207, .299] \\
Daily returns & .253 & .018 & .236 [.200, .269] \\
Frequently updated prices & .226 & .015 & .210 [.178, .250] \\
\bottomrule
\end{tabular}
\end{table}

\begin{figure}[H]
\centering
\includegraphics[width=\linewidth]{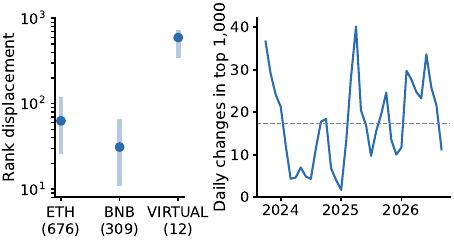}
\caption{Hub prices reorder the \acs{CMC} top 5,000, October 2023 to September 2026. Left: how far a token's rank moves from its rank at its hub's 1 January 2025 price, median and interquartile range (tokens in parentheses). Right: tokens of the top 1,000 replaced each day by hub moves, monthly mean; dashed: overall mean.}
\label{fig:app-rankings}
\end{figure}

\textit{Rankings.} Hub prices also reorder aggregator rankings. In the \ac{CMC} top 5,000 on 1,087 days from October 2023 to September 2026, the median hub-paired token sits 52 ranks from the rank it would hold with its hub's price fixed at its 1 January 2025 level. VIRTUAL-paired tokens move most, by a median of 594 ranks, against 63 for ETH-paired and 31 for BNB-paired tokens (\Cref{fig:app-rankings}, left). Hub moves alone replace 17.3 tokens of the top 1,000 a day on average, and 40 a day in April 2025 (right). The hub price accounts for 26.8\% of the 30-day rank changes of ETH-paired tokens, 37.7\% for VIRTUAL-paired and 19.1\% for BNB-paired tokens.

\subsection{Exit sharing}\label{app:exit}

\textit{Hubs with thin exits.} \Cref{tab:app-hubs} lists the hubs of \Cref{fig:bias2} (left). Among the designated hubs, only VIRTUAL and BNB have less exit capacity than the hub held in their leaves' pools. The nine local hubs whose ratio exceeds 10 have exits of \usd{16}k to \usd{12.5}M, while the tokens priced through them carry \usd{10.7}M to \usd{3.0}bn of market cap. On Base, ten tokens with \usd{367}M of market cap are priced through ELEVATE, whose exit to USD is \usd{16}k and whose leaves' pools hold 305 times that.

\begin{table}[H]
\footnotesize
\caption{Local hubs carry large market caps on thin exits: tokens priced through each hub of \Cref{fig:bias2} (left) and their market cap, the hub's exit capacity to USD and the hub held in its leaves' pools (USD millions), and their ratio, the cut ratio of \Cref{tab:compare}.}
\label{tab:app-hubs}
\begin{tabular}{@{}llrrrrr@{}}
\toprule
Hub & Chain & Tokens & Mkt cap & Exit & In leaf pools & Ratio \\
\midrule
\multicolumn{7}{@{}l}{\emph{Designated hubs}} \\
ETH & Ethereum & 3,317 & 167,054 & 756 & 400 & 0.53 \\
ETH & Base & 1,408 & 43,887 & 434 & 70.0 & 0.16 \\
VIRTUAL & Base & 273 & 271 & 11.4 & 12.1 & 1.07 \\
AERO & Base & 3 & 4.2 & 35.2 & 0.286 & 0.01 \\
ZORA & Base & 24 & 1.5 & 2.1 & 0.416 & 0.20 \\
BNB & BSC & 3,602 & 79,081 & 199 & 288 & 1.45 \\
\multicolumn{7}{@{}l}{\emph{Local hubs, ratio above 10}} \\
ELEVATE & Base & 10 & 367 & 0.016 & 4.7 & 305 \\
VGOLD & BSC & 1 & 119 & 0.020 & 2.3 & 114 \\
Anome & BSC & 1 & 10.7 & 0.077 & 7.0 & 91 \\
GREEN & Base & 2 & 75.5 & 0.016 & 1.3 & 85 \\
ELEVATE & Ethereum & 10 & 1,912 & 0.100 & 4.2 & 42 \\
ADS & Base & 95 & 570 & 0.206 & 8.3 & 40 \\
SENTIS & BSC & 2 & 177 & 1.0 & 35.7 & 36 \\
WIN & Ethereum & 1 & 679 & 0.069 & 1.8 & 26 \\
MGC & BSC & 6 & 3,045 & 12.5 & 137 & 11 \\
\bottomrule
\end{tabular}
\end{table}

\subsection{Value backflow}\label{app:backflow}

\begin{table*}[t]
 \caption{Hub returns rise with leaf inflows before each hub's Binance perpetual listing (VIRTUAL December 2024, ZORA July 2025): same-hour response $b_0$ and six-hour response $\sum b_k$, in \acsp{bp} of hourly return per \acsp{bp} of supply flowing into leaf pools, with $p$-values.}
 \label{tab:backflow-flows}
 \centering
 \begin{tabular}{lcccc}
 \toprule
 & \multicolumn{2}{c}{VIRTUAL} & \multicolumn{2}{c}{ZORA} \\
 \cmidrule(lr){2-3}\cmidrule(lr){4-5}
 Specification & $b_0$ & $\sum b_k$ & $b_0$ & $\sum b_k$ \\
 \midrule
 Main & 38.6 ($<$.001) & 49.9 ($<$.001) & 17.0 (.073) & 42.8 (.028) \\
 Before Binance perpetual & 68.3 ($<$.001) & 86.0 ($<$.001) & 74.7 (.003) & 205.2 ($<$.001) \\
 After Binance perpetual & $-14.4$ (.014) & $-5.7$ (.456) & 8.7 (.265) & 20.4 (.101) \\
 Extreme flows trimmed & 32.3 ($<$.001) & 50.9 ($<$.001) & 46.0 ($<$.001) & 86.1 ($<$.001) \\
 Daily data & 54.9 ($<$.001) & 71.9 ($<$.001) & 52.2 (.012) & 77.3 (.023) \\
 Binance hub return as control, listed hours & 1.10 (.185) & $-1.03$ (.423) & 1.06 (.344) & $-0.31$ (.711) \\
 \bottomrule
 \end{tabular}
\end{table*}

\begin{figure*}[t]
\centering
\includegraphics[width=0.75\linewidth]{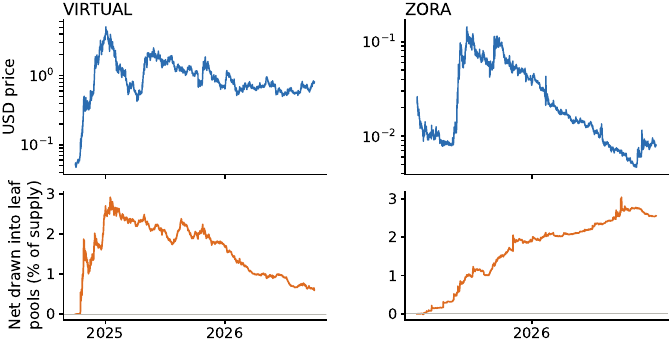}
\caption{Hub prices and the hub drawn into leaf pools on Base: USD price (top, log scale) and the hub its leaves' pools have drawn on net, in percent of supply (bottom).}
\label{fig:app-flows}
\end{figure*}

\textit{Hub prices and leaf flows.} VIRTUAL's price and the hub flowing into its leaves' pools rose together until January 2025, when the leaves' pools had drawn 2.9\% of its supply on net, and fell together afterwards (\Cref{fig:app-flows}). ZORA's leaves kept drawing the hub, up to 3.1\% of its supply in July 2026, while its price fell from its August 2025 peak, after its Binance perpetual listing. The regressions agree (\Cref{tab:backflow-flows}). The relation is strong before each hub's Binance perpetual listing and disappears after it.

\end{document}